\documentclass[10pt]{iopart}

\usepackage{epsfig}
\usepackage{graphics}
\usepackage{graphicx}
\usepackage{bm}
\usepackage{color}
\usepackage{hyperref}
\usepackage{amssymb}
\usepackage{longtable}
\usepackage{rotating}
\usepackage{color}
\usepackage{epsf}
\usepackage{fancyhdr}

\newcommand{\be}{\begin{equation}}
\newcommand{\ee}{\end{equation}}
\newcommand{\ba}{\begin{eqnarray}}
\newcommand{\ea}{\end{eqnarray}}

\def\ltsima{$\; \buildrel < \over \sim \;$}
\def\simlt{\lower.5ex\hbox{\ltsima}}
\def\gtsima{$\; \buildrel > \over \sim \;$}
\def\simgt{\lower.5ex\hbox{\gtsima}}
\begin{document}

\title[Assigning confidence to inspiral GW candidates with model selection]{Assigning confidence to inspiral gravitational wave candidates with Bayesian model selection}

\author{John Veitch and Alberto Vecchio}
\address{School of Physics and Astronomy, University of Birmingham, 
  Edgbaston, Birmingham B15 2TT, UK}
\eads{jveitch@star.sr.bham.ac.uk, av@star.sr.bham.ac.uk}
\date{\today}

\begin{abstract}
Bayesian model selection provides a powerful and mathematically transparent framework to tackle hypothesis testing, such as detection tests of gravitational waves emitted during the coalescence of binary systems using ground-based laser interferometers. Although its implementation is computationally intensive, we have developed an efficient probabilistic algorithm based on a technique known as nested sampling that makes Bayesian model selection applicable to follow-up studies of candidate signals produced by on-going searches of inspiralling compact binaries. We discuss the performance of this approach, in terms of ``false alarm rate'' and ``detection probability'' of restricted second post-Newtonian inspiral waveforms from non-spinning compact objects in binary systems. The results confirm that this approach is a viable tool for detection tests in current searches for gravitational wave signals.
\end{abstract}


\section{Introduction}

Binary systems of black holes and/or neutron stars are one of the most promising sources that are searched for with gravitational wave laser interferometers~\cite{CutlerThorne:2000}, LIGO~\cite{BarishWeiss:1999}, Virgo~\cite{virgo}, GEO-600~\cite{geo} and TAMA-300~\cite{tama}. These instruments have now reached a sensitivity that is astrophysically relevant~\cite{ligoS5}, and near-to-mid future instrumental upgrades are expected to yield further improvements~\cite{upgrades}.
The task of analysing the data from these instruments is computationally intensive due to the weakness of the signals and their large parameter spaces.  The search for gravitational waves from these sources consist of two stages: firstly data segments containing potential candidate signals are identified, and then further scrutinised to decide whether or not gravitational radiation is present.
The latter stage is now the topic of active analysis development and investigation of efficient, robust methods to maximise the observational capabilities of the instruments. A summary of the techniques currently adopted in these follow-up studies for binary inspirals is provided in Ref.\cite{Gouaty-gwdaw12}.


In a previous work we described the application of Bayesian model selection to a simplified
version of this problem, and performed a series of tests to assess its viability~\cite{nest-paper1}.
We introduced a probabilistic algorithm, ``nested sampling''~\cite{Skilling:AIP}, to overcome the difficulty of performing the
multi-dimensional integral, see Equation~(\ref{eqn:Lintegral}), in the case of a 4-parameter Newtonian inspiral signal, with the view of  establishing a basic algorithm to be then 

extended to higher-dimensional models that are necessary to describe in a faithful way inspiral wave-forms.
Specifically, we used Newtonian, equal-mass binary waveforms embedded in stationary Gaussian noise
following roughly the noise spectrum of a first generation gravitational wave interferometer. Having found our
technique promising in the simplest case, we then added un-modelled components to the noise: an instrumental ringdown
signal consisting of a decaying sinusoid and a Poissonian noise component. The intention of these tests was to simulate
the un-modelled artefacts which occur in real interferometer noise. In the first case, the presence of the ringdown
signal did not increase the odds ratio until it reached a signal-to-noise ratio (SNR) of at least 100. In the second case, the
presence of the Poissonian noise component did reduce (as expected) the calculated Bayes factors, but detection was still possible.


In this paper, we present the results of the extension of the application of this algorithm to the family of waveforms that is currently used in all the searches for binary in-spirals from non-spinning compact objects, namely a 5-parameter inspiral waveform approximated to the restricted second post-Newtonian order~\cite{Blanchet-et-al:1996}. This extension is still confined to the case of observations carried out with a single interferometer as an intermediate step towards developing an algorithm that can be applied to an arbitrary number of instruments operating in coincidence. We have also used a more realistic noise spectrum, based on the actual performance of the LIGO Hanford Interferometer during S5, and generated a large number of different synthetic noise realisations from this spectrum to explore the
resulting distribution of Bayes factors and its impact on performance.
With this close approximation to the physical case, we have then performed a multitude of runs on data generated by adding signals with different SNRs to Gaussian and stationary noise to confirm the continued applicability of the algorithm to the more complex case.

The results of the analysis are documented in Section \ref{s:results}, and a short summary of the on-going work and future plans is presented in Section~\ref{s:conclusions}. In the next Section we first explain the theoretical basis of model selection upon which this work rests.

\section{Model Selection}\label{Mod_sel}

\subsection{Models}
The purpose of this analysis is to determine the ratio of probabilities between two possible hypotheses labelled $H_S$ and $H_N$, each of which represents a precise statement about the world. $H_N$, the noise model, corresponds to the statement ``there is no gravitational wave present, and the observed data are drawn from a stationary, Gaussian distribution described by a known noise power spectral density $S_h(f)$.'' $H_S$, the signal model, states ``a gravitational wave described by a known waveform model and parameterised by the vector $\vec{\theta}$ of unknown source parameters is present in the data, \emph{in addition} to the noise described by $H_N$.''

Here we use a five-parameter restricted second post-Newtonian~\cite{Blanchet-et-al:1996} description of the gravitational wave-form in the frequency domain, parameterised by the variables $M =M_1+M_2$, the total mass of the two coalescing objects; $\eta =M_1M_2/(M_1+M_2)^2$, the symmetric mass ratio; $t_c$, the time of coalescence; $\phi_0$, the initial phase of the wave, and $A$ the amplitude of the signal. These five parameters form the parameter vector $\vec{\theta}$, needed to describe a source observed by a single interferometer. The waveform is calculated using the stationary phase approximation in the frequency domain~\cite{Droz-et-al:1999} with the LSC Analysis Library (LAL) functions \texttt{LALInspiralAmplitude} and \texttt{LALInspiralWave}. Further details of the waveform and its implementation in the LAL package are provided in~\cite{LALpage}.

The noise power spectral density of the data is chosen by computing the mean power spectral density $S_h$ using 2\,048 seconds of data from the H1 interferometer with a sampling frequency of 1\,024~Hz, using non-overlapping Tukey windows of length 8 seconds and window parameter $\beta=0.25$. This power spectral density was then used to generate random complex data points $\tilde n(f_k)=\tilde n_k$, each of which had variance $\sigma_k^2 = \frac{1}{2}TS_h(f_k)$, where $T=8\,$s, the length of the segment of observation (in Eqs.(~\ref{eqn:noiseL}) and (~\ref{eqn:Lintegral}) below we indicate with $\Re{\sigma_k}$ and $\Im{\sigma_k}$ the standard deviation of the real and imaginary part of the noise datum $\tilde n_k$). Note the noise spectrum obeys the equation $\sigma_k^2=\Re{\sigma_k}^2+\Im{\sigma_k}^2$. When signals were injected, the data was then given by $\tilde d_k=\tilde n_k+\tilde h_k$. By this method we can use a realistic noise spectrum while ensuring the actual data used is Gaussian and free from any artefacts which may be present in real interferometer data.

The two hypotheses under consideration can be surmised thus,
\begin{itemize}
\item $H_N$: The data is described by $\tilde d_k=\tilde n_k$.
\item $H_S$: The data is described by $\tilde d_k=\tilde n_k+\tilde h_k(\vec{\theta})$.
\end{itemize}

\subsection{Hypothesis testing}

Using Bayes' theorem, the posterior probability for each hypothesis $H$ is given by
\begin{equation}
P(H|\{\tilde d_k\},I)=\frac{P(H|I)P(\{\tilde d_k\}|H,I)}{P(\{\tilde d_k\}|I)}\,,
\end{equation}
and the ratio of probabilities or ``odds ratio'' of the two models is
\begin{eqnarray}
\frac{P(H_S|\{\tilde d_k\},I)}{P(H_N|\{\tilde d_k\},I)}&=&\frac{P(H_S|I)}{P(H_N|I)}\frac{P(\{\tilde d_k\}|H_S,I)}{P(\{\tilde d_k\}|H_N,I)}\\
&=&\frac{P(H_S|I)}{P(H_N|I)}B_{SN}\,,
\end{eqnarray}
the product of the \emph{prior odds ratio} $P(H_S|I)/P(H_N|I)$ and the \emph{Bayes factor} $B_{SN}$, which is the ratio of the marginalised likelihoods of each hypothesis.  In these expressions, the term like $P(H|\{\tilde{d}_k\},I)$ is the posterior model probability, $P(H|I)$ is the prior model probability, $P(\{\tilde{d}_k\}|H,I)$ is the likelihood of the data given the model and $P(\{\tilde{d}_k\}|I)$ is a normalisation constant which cancels in the ratio. We have focussed mainly on the calculation of the Bayes factor in this paper, but some consideration of the prior odds is given in section \ref{ss:priors}.

As the data are drawn from a Gaussian distribution with mean zero, the marginal likelihoods can be written as
\begin{eqnarray}
P(\{\tilde d_k\}|H_N,I)&=&\prod_k\left[2\pi\Re{\sigma_k}\Im{\sigma_k}\right]^{-1}\exp\left(-\frac{\left|\tilde d_k\right|^2}{2\sigma_k^2}\right)\label{eqn:noiseL}\\
P(\{\tilde d_k\}|H_S,I)&=&\int_{\bf \Theta}p(\vec{\theta}|H_S,I)\prod_k\left[2\pi\Re{\sigma_k}\Im{\sigma_k}\right]^{-1} \nonumber\\
&\times&\exp\left(-\frac{\left|\tilde d_k-\tilde h_k(\vec{\theta})\right|^2}{2\sigma_k^2}\right)d\vec{\theta}\label{eqn:Lintegral},
\end{eqnarray}
where $p(\vec{\theta}|H_S,I)$ is the prior probability density distribution on the parameter space.
Equation \ref{eqn:noiseL} is calculated easily since there are no free parameters and no integration is required, however equation \ref{eqn:Lintegral} requires the integration over the parameter space ${\bf \Theta}$, which becomes increasingly difficult as the dimensionality of the parameter space increases.

To overcome this problem, we have employed an algorithm named Nested Sampling, first described by Skilling~\cite{Skilling:AIP}. This allows the calculation of this integral in reduced time by means of a probabilistic approach, whereby the prior distribution on the parameter space is stochastically sampled within a region bounded by constant likelihood. The likelihood bound is iteratively increased as the integration progresses until eventually the entire posterior distribution is covered. A good introduction to this technique can be found in \cite{Sivia}. This same algorithm was successfully used in the four parameter equal-mass Newtonian binary inspiral case~\cite{nest-paper1}; this paper will describe the results obtained with the extension to the model $H_S$ described above, that brings it in line with the requirements for current searches for inspiral binaries from non-spinning compact objects, see \emph{e. g.}~\cite{S3S4inspiral}.

\subsection{Priors}\label{ss:priors}

The prior probability density $p(\vec{\theta}|H_S,I)$ used was uniform in all parameters within the range: $1\le{}A/A_0\le{}30$ (where $A_0$ corresponds to a source lying directly above the interferometer at a distance of 1 Mpc); $0\le{}\eta\le{}0.25$, the entire possible range; $0\le\phi_0<2\pi$; $-0.005~{\rm s}\le{}t_c-t_\mathrm{inj}\le{}0.005~{\rm s}$; $0.8M_\mathrm{inj}\le{}M\le{}1.2M_\mathrm{inj}$. $t_\mathrm{inj}$ and $M_\mathrm{inj}$ are the time of coalescence and the total mass of the signal that was super-imposed to Gaussian, stationary noise (so-called ``software injections''). The SNR was varied by changing the amplitude $A$, holding the total mass constant.
The value of total mass for the injected waveform was chosen so as to keep the innermost stable circular orbital frequency
below 512\,Hz, allowing us to decrease the computational expense by making the data above this frequency irrelevant to the calculation. The mass parameter values were set to $M_\mathrm{inj}=11.6\,M_{\odot}$ and $\eta_\mathrm{inj}=0.2107$, corresponding to component masses of 8.1 and 3.5 solar masses.

\subsection{Prior odds ratio}\label{ss:priorodds}

The prior odds ratio $P(H_S|I)/P(H_N|I)$, is a factor which describes the ratio of probabilities assigned to each hypothesis, prior to the information from the data set $\{\tilde d_k\}$ being analysed. It quantifies our prior belief (or disbelief) in the relative probabilities of the models. Since we expect this to be $\ll{}1$, it acts as a threshold value of odds which must be overcome to render the total posterior odds $>1$, and provide a greater net belief in the signal hypothesis. The final result of the analysis is not a statement of detection or not, but the quantified ratio of beliefs in the two situations. If the final odds are large enough to overcome the threshold and it is believed that the data is adequately described by the models, then a claim of detection could be considered. It is to be emphasised that the analysis itself cannot make this decision, merely provide the best information on which to base it. As we shall see in Section \ref{ss:signals}, Bayesian model selection can provide virtual certainty, but not actual certainty. 

The question of the prior odds is not conclusively answered here, but we shall demonstrate how they can be chosen. In fact, the posterior odds cannot be computed without knowing the prior odds. From a strictly Bayesian point of view, one would estimate the probability of observing an inspiral signal with parameters within the prior ranges, based on the knowledge of the rate of such events from other sources, such as theoretical considerations, studies of the population of progenitor systems, and current observations (in other observational windows) of binary systems~\cite{rates}. In practice, our knowledge on these parameters is likely too poor to be used in practice.

An alternative, and probably much more viable approach -- in particular for applications on real data where the statistical properties of the noise are not exactly known -- is to consider the "false alarm rate" of posterior odds which in turn can be used to set a particular prior odds ratio to use. In order to do this we must also reduce the assignment of continuous probabilities to a binary choice between detection and non-detection. This condensation eliminates some of the additional information that the odds ratio contains, such as the confidence of detection, but is useful for considering what threshold values are appropriate in this approach.

If we re-write the posterior odds ratio using shorthand $P(H_N|I)/P(H_S|I)=T_{SN}$:
\begin{equation}
\frac{P(H_S|\{d_k\},I)}{P(H_N|\{d_k\},I)} = \frac{B_{SN}}{T_{SN}}
\end{equation}
we can {\em interpret} the quantity $T_{SN}$ -- the inverse of the prior odds ratio -- as a threshold on the Bayes factor that needs to be overcome to produce a posterior odds ratio larger than a given value. If we arbitrarily designate detection as the case when the posterior odds ratio is, say greater than unity, then we can define the false alarm rate $F(T_{SN})$ as the number of trials on a noise-only data set which produce $P(H_S|\{d_k\},I)/P(H_N|\{d_k\},I)>1$, divided by the total number of trials $M$ for a given value of $T_{SN}$:
\begin{equation}
F(T_{SN})=M^{-1}\sum_i^M\left\{
\begin{array}{cc}
1 &: B_{SN}>T_{SN}\\
0 &: B_{SN}\le{}T_{SN}
\end{array}\right.
\label{e:TSN}
\end{equation}
Note that the threshold quantity $T_{SN}$ is defined so that larger thresholds produce lower false alarm rates. In Section \ref{ss:nosig} we will show how the prior odds ratio can be chosen to give a desired ``false alarm rate'' when the code is run repeatedly over different noise realisations. By tuning (the inverse of) the prior odds ratio, i.e. the threshold $T_{SN}$, one can control both the ``false alarm'' and ``false dismissal'' rates, and therefore design an algorithm that perform with a given detection efficiency as we will show in Section~\ref{ss:signals}.

\section{Results}
\label{s:results}
\subsection{Synthetic noise with no signal}\label{ss:nosig}

This section deals with the distribution of Bayes factors recovered by the algorithm when no signal is present. In this case, the data is described entirely by the stationary, Gaussian noise $\tilde d_k=\tilde n_k$. However, since $\tilde n_k$ is drawn from a distribution, each individual noise realisation will produce a different result. 

\begin{figure}
\begin{center}
\resizebox{\columnwidth}{!}{\includegraphics{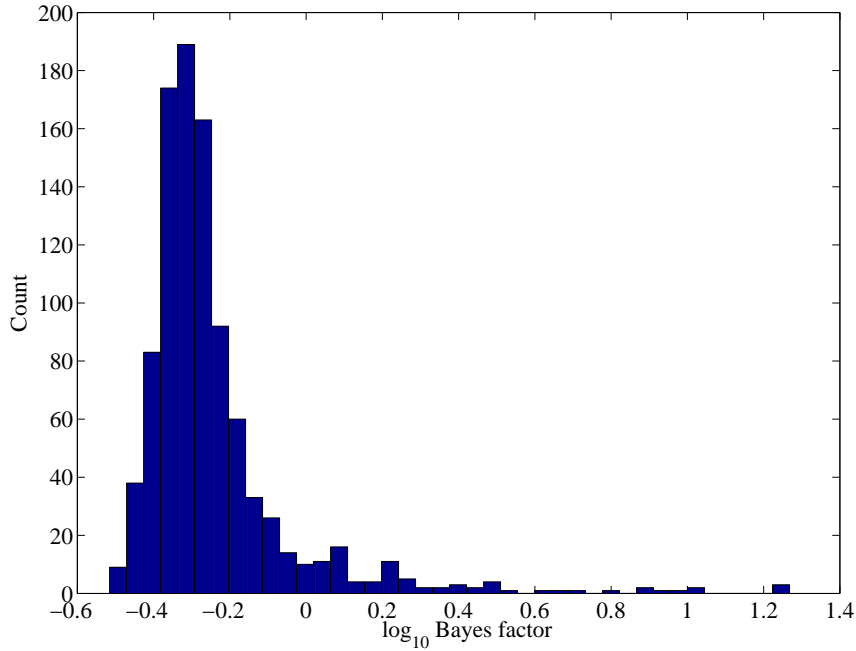}}
\caption{\label{fig:nulldist}The distribution of Bayes factors $B_{SN}$ when the algorithm is applied to stationary, Gaussian noise as described in Section \ref{Mod_sel} with no added signal. 969 trials are performed drawing data from a noise spectral density that is representative of the LIGO Hanford 4-km instrument during the fifth science run. The distribution clearly shows that in the majority of cases, the data produces a Bayes factor less than 1, meaning that the data lends stronger support to the noise model than the signal model. However, the distribution has a long tail above $B_{SN}=1$, which will influence the choice of prior odds.}
\end{center}
\end{figure}

In Figure \ref{fig:nulldist}, we show the results of running the algorithm with the prior ranges as described in section \ref{ss:priors} a total of $M=969$ times, see Eq.~\ref{e:TSN}. The time to complete one run of the code is variable due to the probabilistic nature of the algorithm, but the approximate mean time to complete one of the 969 jobs was 4 hours 45 minutes on a single 2.6\,GHz Opteron CPU. When compared to the run-time of a Reversible Jump MCMC algorithm which also computes an approximation to the Bayes factor for a similar problem, reported in \cite{RichardMCMC} to be several days on a cluster of 1\,024 Xeon processors, this demonstrates the efficiency of this method.
Only the distribution of Bayes factors are shown, and we can see that placing a threshold $T_{SN}=1,$ which corresponds to equal belief in the two hypotheses $H_S$ and $H_N$ would produce a false alarm rate of $F=80/969=8.3\,\%$.
This false alarm rate seems unacceptably high, but that is because it is based on an unreasonable initial assumption, that inspiral signals with parameters lying within the prior range are expected as often as they are not. Given that our prior odds are $50\%$, the use of the Bayes factor has shown us that in 91.7\% of cases, the data has shifted the balance of probabilities to favour the noise model. This is reassuring, since there was no signal injected.

If we are to change our prior odds, or threshold value to, say, $T_{SN}=10$, a statement that we find $H_N$ only ten times as probable as $H_S$, we have reduced the false alarm rate to $F(10)\approx0.5\%$.

It should be remembered that the results shown here are obtained using Gaussian stationary noise, whereas in a real interferometer the data also
contains artefacts of various types. Although our previous work showed the algorithm to be robust against some such disturbances~\cite{nest-paper1}, we do not know the effect of such disturbances on the false alarm rate in the case of considering only two models.
We believe it is likely to increase the false alarm rate, and therefore the required threshold, and such an evaluation would have to be done as part of the analysis of science data using this method. In order to eliminate the (extremely small) background of true gravitational waves, data from multiple interferometers could be combined incoherently, while using a multiple-interferometer version of the method presented here, which translates in a straightforward change of the likelihood function; this is a procedure similar to the use of time-slides in the current analyses on real data~\cite{S3S4inspiral} and we are currently developing the necessary software to carry out such studies. We should also not discount the consideration of astrophysical rates of observable events~\cite{rates}, as it seems to us that such an estimation is likely to produce a higher threshold than the consideration of false alarms (using this algorithm). 

In addition to controlling the rate of ``false alarms'', the prior odds ratio controls the rate of ``false dismissal'', by a means that we shall investigate in the next section.

\subsection{Synthetic noise with a variety of signal strength}\label{ss:signals}

Here we describe the results obtained when injecting signals to Gaussian and stationary noise at a range of signal to noise ratios (SNRs) and running the algorithm to find the Bayes factor. Unlike our earlier study~\cite{nest-paper1}, we are now interested in the range of results which occur when different noise realisations are present. To this end, at each value of SNR, we used fifty different noise realisations and repeated the test on each of them (the choice of the number of trials was  simply determined by computational constraints and we plan to carry out a much more systematic investigation in the near future). This allowed us to determine the mean and distribution of the Bayes factor as a function of SNR, defined as ${\mathrm SNR}=\sqrt{\sum|d_k|^2\sigma_k^{-2}}$.
In each case, the parameters of the injected signal were the same, save the amplitude, and were the same as described in Section \ref{ss:priors}.

\begin{figure}
\begin{center}
\resizebox{\columnwidth}{!}{\includegraphics{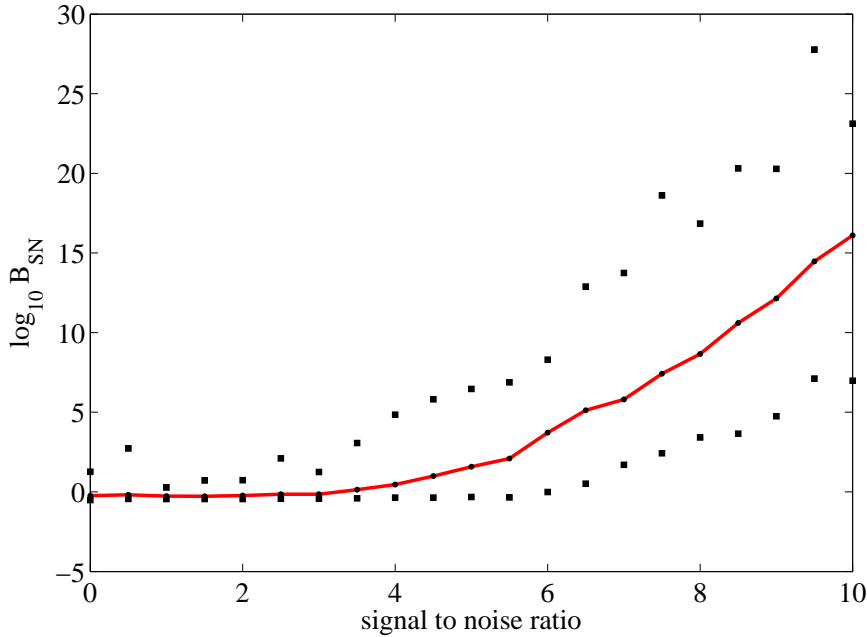}}
\caption{\label{fig:diffnoise} Results of running the algorithm on a range of signals with different signal-to-noise ratios. For each SNR, 50 different noise realisations were used to find the distribution of the Bayes factor. The plot shows the steady increase of the mean log Bayes factor (solid), and the minimum and maximum recovered values (squares).}
\end{center}
\end{figure}

Figure \ref{fig:diffnoise} shows the results of this procedure at SNR intervals of 0.5 between values of zero and ten. Note the logarithmic scale, indicating the extremely rapid growth of the Bayes factor as SNR increases. However, the range of values found also increases. Examining the minimum recovered values shows that there are some noise realisations which produce a Bayes factor less than 1 up until SNR 6. The maximum values fluctuate due to the small numbers of trials performed, but show a general increase, and also that the range upwards is larger than that downwards, in accordance with the distribution in Figure \ref{fig:diffnoise}.

\begin{figure}
\begin{center}
\resizebox{\columnwidth}{!}{\includegraphics{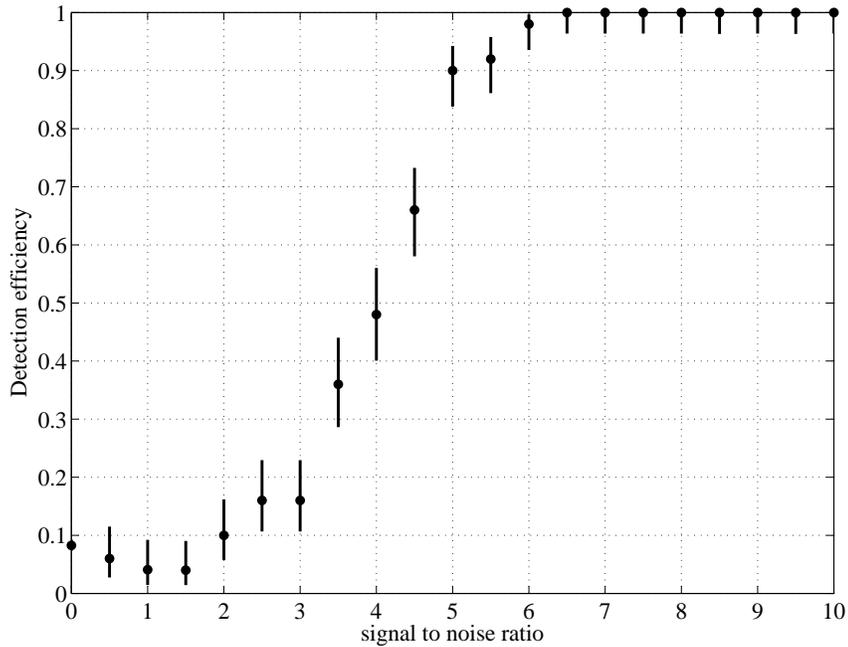}}
\caption{\label{fig:eff1}The detection efficiency, defined as the number of detected signals divided by the number of trials performed with signals at a given SNR, for $T_{SN}=1$, see Eq~(\ref{e:TSN}). Here detection is arbitrarily defined as the case in which the posterior odds ratio satisfies the condition $P(H_S|\{d_k\},I)/P(H_N|\{d_k\},I) > 1$.}
\end{center}
\end{figure}

\begin{figure}
\begin{center}
\resizebox{\columnwidth}{!}{\includegraphics{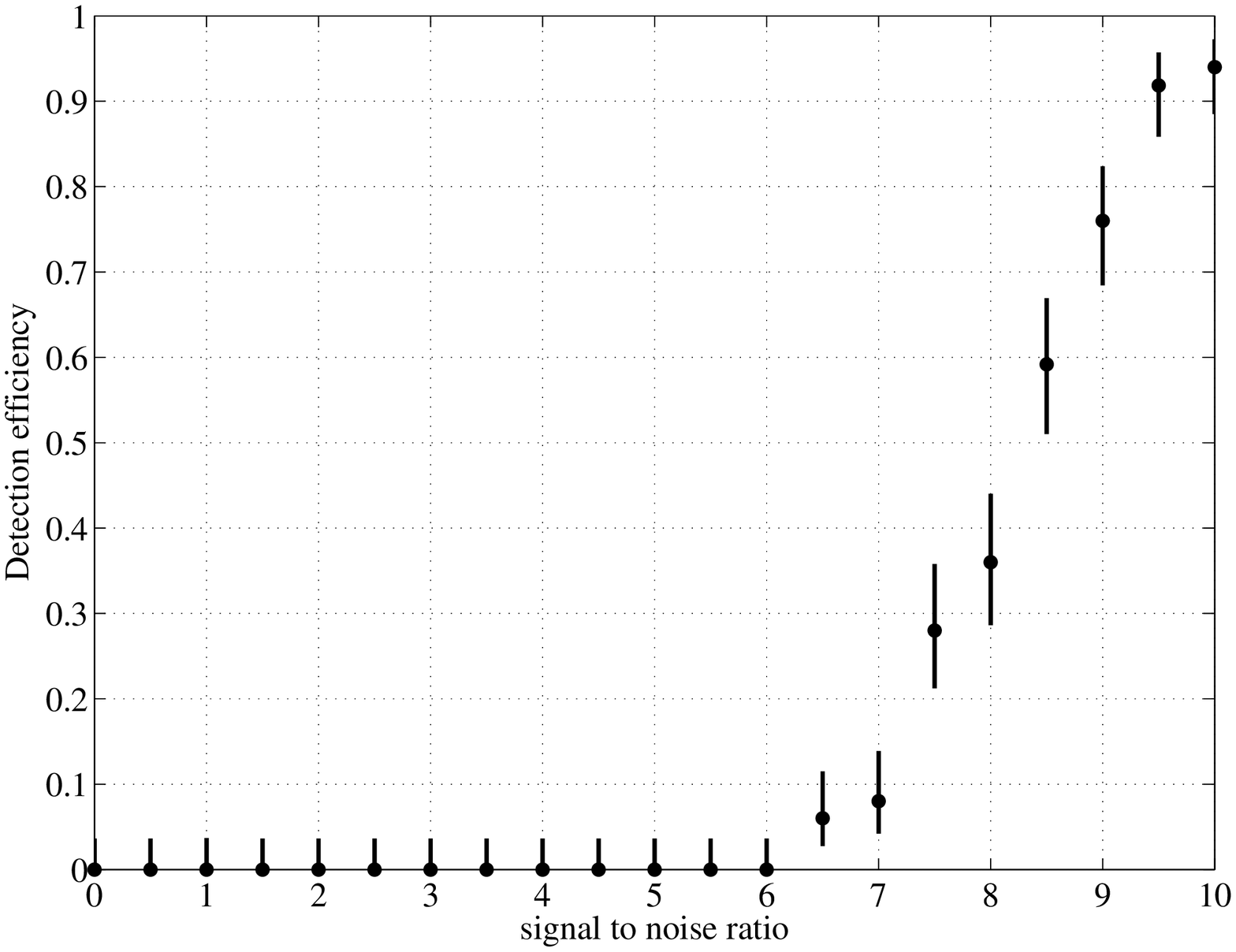}}
\caption{\label{fig:eff2}Detection efficiency for $T_{SN}=3.15\times10^9$. Same as Figure~\ref{fig:eff1}, but with a different choice of the prior odds ratio.}
\end{center}
\end{figure}

We can now see the effect of changing the prior odds ratio on the posterior probability: it will change the point at which the Bayes factor is sufficient to tilt the odds in favour of the signal model. In Figures \ref{fig:eff1} and \ref{fig:eff2} we have plotted the detection efficiency curve using the same criterion for ``detection'' as in Section \ref{ss:nosig}, namely a posterior odds ratio greater than one. In Figure \ref{fig:eff1}, the prior odds or threshold is set to $T_{SN}=1$, assigning equal probabilities to both hypotheses \emph{a-priori}. The detection efficiency, defined as the number of detected injections divided by the number of injections performed at that SNR, undergoes a sharp transition between SNR 3 and 6, above which detection is almost certain. However, the false alarm rate is high, since we are using $T_{SN}=1$, and this can be seen in the non-zero detection efficiency at SNR$ = 0$. The sensitivity available at this prior odds is reduced by decreasing our prior probability in the signal model, but the same must be true of any analysis that does not assign equal probabilities to $H_S$ and $H_N$. In the framework of Bayesian inference we have direct control over this parameter and can readily see its effect.

As a further example, we have performed the same efficiency calculation, but this time using a value for the prior odds ratio driven by (overoptimistic) rate estimates. Let us assume that we expected to see one inspiral per year within the prior range of mass and distance stated in Section \ref{ss:priors}; in the 10\,ms prior range of $t_c$, the rate is equal to a prior odds ratio of $3.17\times{}10^{-10}$, or $T_{SN}=3.15\times{}10^9$. This much increased threshold needs a signal with a higher SNR in order to achieve a correspondingly higher Bayes factor. In fact even with SNR = 10 detection is only $\sim{}95\%$ certain. The false alarm rate at this threshold is too low to calculate for these examples.

We wish to emphasise again that the rate figure used in this example is purely for illustration, see~\cite{rates}, but it shows that reasonable disbelief in the signal hypothesis increases the standard of evidence needed to claim detection. The same parameter, the prior odds, tunes both false alarm and false dismissal rates.

\section{Conclusions}
\label{s:conclusions}

We have shown how the conceptually simple framework of Bayesian hypothesis testing and its computationally efficient implementation using a ``nested sampling'' algorithm which we described in~\cite{nest-paper1} can be naturally extended to the realistic case of follow-up studies of binary inspirals modelled at the second post-Newtonian order. 
There is now growing evidence that the computational expense which has discouraged the use of Bayesian hypothesis testing in the past can be ameliorated by the use of probabilistic algorithms such as nested sampling~\cite{Skilling:AIP,nest-paper1} or 
reversible jump~\cite{RichardMCMC,CornishLittenburg} and/or delayed rejection~\cite{GreenMira} Markov Chain Monte Carlo methods. These algorithm provide viable practical implementations of Bayesian model selection for on-going gravitational wave searches and as they reach maturity their performance relative to existing follow-on approaches~\cite{Gouaty-gwdaw12} can be tested in order to assess the relative merits of the different strategies.

We are currently working on an extension of the algorithm presented in this paper to observations with multiple interferometers and on a more extensive characterisation of its performance on real data.

\section*{Acknowledgments}

This work has benefited from many discussions with members of the LIGO Scientific Collaboration. The computations were performed on the Blue BEAR and Tsunami Beowulf clusters of the University of Birmingham. This work has been supported by the UK Science and Technology Facilities Council.

\end{document}